\documentclass[preprintnumbers,article,amsmath,amssymb,floatfix,10pt,prd,onecolumn,
superscriptaddress,nofootinbib]{revtex4}
\usepackage{bbm}
\usepackage{amsfonts}
\usepackage{mathrsfs}
\usepackage{latexsym}
\usepackage{epstopdf}
\usepackage{epsfig}
\usepackage{epstopdf}
\usepackage{graphicx}
\usepackage{amssymb}
\usepackage{amsmath}
\usepackage{dcolumn}
\usepackage{bm}
\usepackage{color}
\usepackage{comment}
\usepackage{xcolor}
\begin{document}

\title{\bf $f(\mathcal{R},\varphi,\chi)$ Cosmology with Noether Symmetry }

\author{M. Farasat Shamir}
\email{farasat.shamir@nu.edu.pk}\affiliation{National University of Computer and
Emerging Sciences,\\ Lahore Campus, Pakistan.}

\begin{abstract}
This paper is devoted to explore modified $f(\mathcal{R})$ theories of gravity using Noether symmetry approach. For this purpose, Friedmann-Robertson-Walker spacetime is chosen to investigate the cosmic evolution. The study is mainly divided into two parts: Firstly Noether symmetries of metric $f(\mathcal{R})$ gravity are revisited and some new class of solutions with the help of conserved quantities are reported.
It is shown that different scenarios of cosmic evolution can be discussed using Noether symmetries and one of the case indicates the chances for the existence of Big Rip singularity. Secondly, $f(\mathcal{R})$ theory coupled with scalar field has been discussed in detail. The Noether equations of modified gravity are reported with three subcases for flat Friedmann-Robertson-Walker universe. It is concluded that conserved quantities are quite helpful to find some important exact solutions in the cosmological contexts. Moreover, the scalar field involved in the modified gravity plays a vital role in the cosmic evolution and an accelerated expansion phase can be observed for some suitable choices of $f(\mathcal{R},\varphi,\chi)$ gravity models.
\end{abstract}

\maketitle

\date{\today}
\section{Introduction}

The accelerated expansion of universe and modified theories of gravity have been two heavily debated topic of discussions in the last two decades. It has been argued that the mysteries like dark energy and dark matter, initial singularity problem and flatness issues can be well addressed in the context of modified or alternative theories of gravity. Based upon original theory of general relativity (GR), a number of modifications have been proposed by constructing intricate Lagrangians. The most discussed and viable theories include $f(\mathcal{R})$ theory of gravity. Nojiri and Odintsov \cite{NJO} are among the pioneers who discussed the possible coupling of matter with curvature.
Some review papers can be really helpful to understand the viable features of $f(\mathcal{R})$ gravity \cite{Fel}-\cite{Fel2}. Further modifications like $f(\mathcal{R},T)$ gravity \cite{HRKO} and a recently proposed $f(\mathcal{R},\varphi,\chi)$ theory of gravity \cite{sabestian} are also among successful theories of gravity. It is expected that these modified theories of gravity may well address the issues of late-time cosmic acceleration using some specific choices of cosmological models.

Noether symmetry approach is the most elegant and systematic way to compute conserved
quantities \cite{IV}. These symmetries smartly minimize the complexities involved in a system of non-linear partial differential equations (PDEs) and many new solutions can be constructed using conserved quantities.
In fact, the conservation laws play an important role in studying different physical phenomenon. The integrability
of PDEs depends on the number of conservation laws.
According to Noether's theorem, any differentiable symmetry of the
action for a physical system corresponds to some conservation law.
This theorem is very important as it provides the information
about the conservation laws in physical theories including GR.
According to Noether theorem \cite{Noethertheorem1}, the translational and rotational symmetries of any object are the consequence of the conservation laws of linear and angular momentum.
Many authors have used this theorem in recent years to discuss some important issues in different cosmological contexts.

Sharif and Waheed \cite{Sharif13} computed the energy contents of stringy charged black hole solutions with the help of approximate symmetries. Kucukakca \cite{Kucuka16} used Noether symmetries to investigate the exact solutions of Bianchi type-$I$ spacetime. The exact solutions in $f(\mathcal{R})$ gravity have been explored using Noether symmetries methods for Friedmann-Robertson-Walker (FRW) spacetime \cite{Cap21}. Fazlollahi \cite{Fazlollahi} used Noether gauge symmetries to obtain an effective equation of state parameter for corresponding cosmology and it was concluded that the model provided viable cosmic scale factor with respect to observational data.  In a recent paper, Bahamonde et al. \cite{Bahamonde} provided a class of new exact spherically symmetric solutions in the context of $f(\mathcal{R},\varphi,\chi)$ theory using Noether's symmetry approach. In another paper \cite{Bahamonde1}, teleparallel gravity models have been studied by adopting the Noether symmetry approach and some exact solutions are derived in the context of flat FRW cosmology. Shamir and Ahmad \cite{Sir&M,Sir&Me2} used Noether symmetries to investigate
the exact solutions of the field equations in $f(\mathcal{G},T)$
theory of gravity and discussed some cosmologically important $f(\mathcal{G},T)$ gravity models with both isotropic and
anisotropic backgrounds. It was reported that
specific models of modified Gauss-Bonnet gravity may be used to
reconstruct $\Lambda$CDM cosmology without involving any
cosmological constant. Thus it seems interesting to use Noether symmetries to further explore the universe with a hope of some fruitful results.

In this paper, we are focused to investigate $f(\mathcal{R},\varphi,\chi)$ gravity using Noether symmetry approach. We choose the flat FRW spacetime for this purpose. The paper is organized in the following way: Some basics of $f(\mathcal{R},\varphi,\chi)$ gravity are given in section \textbf{2}. Section \textbf{3} provides a detailed discussion about  symmetry reduced Lagrangian and Noether equations for FRW universe model in $f(\mathcal{R},\varphi,\chi)$ gravity. Cosmological solutions based upon conserved quantities are presented in section \textbf{4}. Last section gives a brief summary of the results.

\section{Some Basics of $f(\mathcal{R},\varphi,\chi)$ Gravity}

The general action for  $f(\mathcal{R},\varphi,\chi)$ gravity is given by \cite{sabestian},
 \begin{equation}\label{action}
\mathcal{S}= \int d^{4}x
\sqrt{-g}\big[\frac{1}{2{\kappa}^{2}}f(\mathcal{R},\varphi,\chi)+\mathcal{L}_{m}\big],
\end{equation}
where $\mathcal{L}_{m}$ stands for usual matter Lagrangian, ${\kappa}^{2}=8\pi G$ and
\begin{itemize}

\item $\mathcal{R}$ is the Ricci Scalar and $\varphi$ is the scalar field,
\item $\chi=-\frac{\epsilon}{2}{\partial}^u\varphi{\partial}_u\varphi$, $\epsilon$ being a parameter such that when equal to $1$ represents canonical scalar field and when equal to $-1$ represents a phantom scalar field.
\item Since $f(\mathcal{R},\varphi, \chi)$ is a multivariate analytic function and its partial derivatives will be involved in many equations in this paper, so for simplicity we use the notations $f(\mathcal{R},\varphi, \chi)\equiv f$, $f_{\mathcal{R}}\equiv\frac{\partial f}{\partial{\mathcal{R}}}$, $f_{\varphi}\equiv\frac{\partial f}{\partial{\varphi}}$ and $f_\chi\equiv\frac{\partial f}{\partial{\chi}}$.
\end{itemize}

It is evident from the action that $f(\mathcal{R},\varphi,\chi)$ gravity contains two additional scalar degrees of freedom. Thus the chances of admitted solutions increase due to more degrees of freedom as compared to usual GR. This makes the theory more interesting as it falls into a rather different class than those modified gravity theories typically considered in the literature. Modified field equations are obtained by varying the action $S$ in Eq.(\ref{action}) with respect to the
metric tensor
\begin{equation}\label{301}
f_{\mathcal{R}}\mathcal{R}_{\mu\nu}-\frac{1}{2}f g_{\mu\nu}-\nabla_{\mu}\nabla_{\nu}f_\mathcal{R}+g_{\mu\nu}\nabla_{\alpha}\nabla^{\alpha}f_\mathcal{R}-\frac{\epsilon}{2}f_\chi(\nabla_{\mu}\varphi)( \nabla_{\nu}\varphi)=\kappa^2 T^m_{\mu\nu},
\end{equation}
where $\Box\equiv\nabla^{\mu}\nabla_{\mu}$ and $T^m_{\mu\nu}$ is the standard energy-momentum tensor.
It is interesting to notice that these field equations are the fourth order PDEs due to the involvement of covariant derivatives and reduce to ordinary $f(\mathcal{R})$ gravity equations if we consider $f(\mathcal{R},\varphi,\chi)=f(\mathcal{R})$ and GR equations when $f(\mathcal{R},\varphi,\chi)=\mathcal{R}$. Moreover, variation of action (\ref{action}) with respect to the scalar field $\varphi$ yields
\begin{equation}\label{4}
\nabla_{\mu}(f_\chi\nabla^{\mu}\varphi)+\epsilon f_{\varphi}=0,
\end{equation}
which is basically the Klein-Gordon equation. It is worth mentioning that the field
equations (\ref{301}) can be rearranged in a form which seems familiar with
usual GR field equations
\begin{equation}\label{402}
G_{\mu\nu}=\mathcal{R}_{\mu\nu}-\frac{1}{2}g_{\mu\nu}\mathcal{R}=T^c_{\mu\nu}+\tilde{T}^m_{\mu\nu},
\end{equation}
where $\tilde{T}^m_{\mu\nu}=T^m_{\mu\nu}/f_\mathcal{R}$ and
energy-momentum tensor for gravitational fluid is given by
\begin{equation}\label{5}
T^c_{\mu\nu}=\frac{1}{f_\mathcal{R}}\big[\frac{1}{2}g_{\mu\nu}\big(f-\mathcal{R}f_\mathcal{R}\big)+\nabla_{\mu}\nabla_{\nu}f_\mathcal{R}-
g_{\mu\nu}\nabla_{\alpha}\nabla^{\alpha}f_\mathcal{R}+\frac{\epsilon}{2}f_\chi\big(\nabla_{\mu}\varphi\big)\big(\nabla_{\nu}\varphi\big)\big].
\end{equation}
It can be seen from Eq.(\ref{402}) that energy-momentum tensor for
gravitational fluid $T^c_{\mu\nu}$ provides matter contents from
geometric origin. Thus, it seems interesting as this  approach may provide
all the matter components which may be essential to unveil the dark
mysteries of universe. Moreover, contracting the field equations Eq.(\ref{301}), it follows that
\begin{equation}\label{502}
\mathcal{R}f_\mathcal{R}+\chi f_\chi-2f+3\Box f_\mathcal{R}=\kappa^2 T^m.
\end{equation}
It is obvious from Eq.(\ref{502}) that $T^m=0$ no longer implies $\mathcal{R}=0$ as in the case of GR. Thus this theory may admit more exact or numerical solutions of the modified field equations. Keeping in view all above mentioned interesting facts of $f(\mathcal{R},\varphi,\chi)$ theory of
gravity, it is hoped to obtain some fruitful results to explore the dark energy issues and phenomenon of
cosmic expansion.

For the present analysis, we choose the FRW space-time
\begin{equation}\label{6}
d{s}^{2}=-dt^{2}+a^2(t)[d{x}^{2}+d{y}^{2}+d{z}^{2}].
\end{equation}
The dynamical quantities ${\mathcal{R}}$ and ${\chi}$ for this spacetime turn out to be
\begin{equation}\label{Ricci}
{\mathcal{R}}=6\big(\frac{\dot{a}^{2}+a\ddot{a}}{a^{2}}\big),~~~~~~~{\chi}=\frac{\epsilon}{2}\dot{\varphi}^2.
\end{equation}
The corresponding field equations and the Klein-Gordon equation for vacuum case take the form
\begin{equation}\label{8}
3\frac{\ddot{a}}{a}f_{\mathcal{R}}+\frac{1}{2}f-3\frac{\dot{a}}{a}\dot{f_\mathcal{R}}+\frac{1}{2}f_\chi (\dot{\varphi})^2=0,~~~~~~
[\frac{\ddot{a}}{a}+2(\frac{\dot{a}}{a})^{2}]f_{\mathcal{R}}+\frac{1}{2}f-2\frac{\dot{a}}{a}\dot{f_\mathcal{R}}-\ddot{f_\mathcal{R}}=0,
\end{equation}
\begin{equation}\label{9a}
\dot{f_\chi}\dot{\varphi}+f_\chi[\ddot{\varphi}+3\frac{\dot{a}}{a}\dot{\varphi}]+f_\varphi=0,
\end{equation}
where overdot is to denote the time derivative.
The advantages of exact solutions in modified theories of gravity have gained much importance and popularity over the recent two decades, especially to explore the phenomenon of cosmic expansion and phase transitions. The analysis of Eqs. (\ref{8})-(\ref{9a}) is quite difficult task as these equations are highly non-linear and complicated in nature due the involvement of multivariate function $f(\mathcal{R},\varphi,\chi)$ and its derivatives. So, there are two choices: One is to solve these equations by using some appropriate numerical or analytical techniques and imposing some physical assumptions. The second possibility is to investigate the theory by using Noether symmetry approach.  In fact, using Noether symmetries some viable cosmological models can be investigated and conserved quantities can be utilized to reconstruct some physically important exact or numerical solutions. Hence the latter approach seems much more interesting and we follow the same in this paper.

\section{Symmetry Reduced Lagrangian and Noether Equations}

Noether symmetries have become quite essential practice to investigate the solutions of non-linear differential equations. In this section, we develop the point-like Lagrangian for FRW spacetime in the context of $f(\mathcal{R},\varphi,\chi)$ gravity and apply Noether symmetry approach to find the corresponding determining equations. The existence of this approach implies the uniqueness of the vector field in the associated tangent space. Thus, the vector field acts like symmetry generator which further provides the conserved quantities helpful in exploring the exact solutions of modified field equations.

We can re-write the action (\ref{action}) in its canonical form in such a way that the number of degrees of freedom are reduced. Thus we have
\begin{equation}\label{32_eqn}
\mathcal{S}=\int dt~\mathcal{L}(a,\dot{a},\mathcal{R},\dot{\mathcal{R}},\varphi,\dot{\varphi}).
\end{equation}
Since $\chi$ depends on $\varphi$, so this action for FRW spacetime takes the form
\begin{equation}\label{32_eqn}
\mathcal{S}=\int dt~a^3[f -\mu_{1}(\mathcal{R}-{\bar{\mathcal{R}}})-\mu_{2}({\chi}-\bar{\chi})].
\end{equation}
This lagrange multiplier arrangement of the action is justified as the constraint equations $\mathcal{R}-{\bar{\mathcal{R}}}=0$ and ${\chi}-\bar{\chi}=0$ provide the actual form of action (\ref{action}) for FRW spacetime.
The Lagrange multipliers $\mu_{1}$ and $\mu_{2}$ after varying with respect to $\mathcal{R}$ and $\chi$ turn out to be
\begin{equation}
\mu_{1}=f_\mathcal{R},~~~~~
\mu_{2}=f_\chi,
\end{equation}
respectively. So Eq.(\ref{32_eqn}) becomes
\begin{equation}\label{302_eqn}
\mathcal{S}=\int dt~a^3[f -f_\mathcal{R}(\mathcal{R}-6\frac{\dot{a}^{2}}{a^{2}}-6\frac{\ddot{a}}{a})-f_\chi({\chi}-\frac{\epsilon}{2}\dot{\varphi}^2)].
\end{equation}
After integrating by parts and ignoring the boundary terms, the point-like Lagrangian takes the following form
\begin{eqnarray}\label{34_eqn}
\mathcal{L}(a,\dot{a},\mathcal{R},\dot{\mathcal{R}},\chi,\dot{\chi})=
a^{3}(f-\mathcal{R}f_{\mathcal{R}}-\chi f_\chi)-6(a\dot{a}^2f_\mathcal{R}+\dot{a}\dot{\mathcal{R}}a^2f_{\mathcal{R}\mathcal{R}})-
6a^2\dot{a}(\dot{\varphi}f_{\mathcal{R}\varphi}+\dot{\chi}f_{\mathcal{R}\chi})+\frac{\epsilon}{2}a^3\dot{\varphi}^2f_\chi.
\end{eqnarray}
It is interesting to notice that when $f(\mathcal{R},\varphi,\chi)=f(\mathcal{R})$, this Lagrangian reduces to the same as investigated for usual $f(\mathcal{R})$ gravity \cite{22,22a},
The Euler-Lagrange equations are given by
\begin{equation}\label{29_eqn}
\frac{\partial \mathcal{L}}{\partial
u^{\mu}}-\frac{d}{dt}\Bigg(\frac{\partial \mathcal{L}}{\partial
\dot{u}^{\mu}}\Bigg)=0.
\end{equation}
For FRW spacetime (\ref{6}) and Lagrangian (\ref{34_eqn}), the Euler-Lagrangian equations turn out to be
\begin{eqnarray}\label{35_eqn}
&&\big(\frac{\dot{a}^2}{a^2}+2\frac{\ddot{a}}{a}\big)f_\mathcal{R}+2\frac{\dot{a}}{a}\dot{f_\mathcal{R}}+\ddot{f_\mathcal{R}}+\frac{1}{2}(f-\mathcal{R}f_{\mathcal{R}}-\chi f_\chi)+\frac{\epsilon}{4}\dot{\varphi}^2f_\chi=0,\\
&&\big[\mathcal{R}-6\big(\frac{\dot{a}^2}{a^2}+\frac{\ddot{a}}{a}\big)\big]f_{\mathcal{R}\mathcal{R}}+\big[\chi-\frac{\epsilon}{2}\dot{\varphi}^2\big]f_{\mathcal{R}\chi}=0,~~~
\big[\mathcal{R}-6\big(\frac{\dot{a}^2}{a^2}+\frac{\ddot{a}}{a}\big)\big]f_{\mathcal{R}\chi}+\big[\chi-\frac{\epsilon}{2}\dot{\varphi}^2\big]f_{\chi\chi}=0.
\end{eqnarray}
The Hamiltonian of the Lagrangian also known as energy functional $E_{\mathcal{L}}$ is given by
\begin{equation}
E_{\mathcal{L}}=\dot{u}^\mu\frac{\partial \mathcal{L}}{\partial \dot{u}^{\mu}}-\mathcal{L},
\end{equation}
which in our case simplifies to
\begin{equation}\label{37_eqn}
E_{\mathcal{L}}=-6a\dot{a}^2f_\mathcal{R}-6a^2\dot{a}(\dot{\mathcal{R}}f_{\mathcal{R}\mathcal{R}}+\dot{\chi}f_{\mathcal{R}\chi})-a^{3}(f-\mathcal{R}f_{\mathcal{R}}-\chi f_\chi)-\frac{\epsilon}{2}a^3\dot{\varphi}^2f_\chi.
\end{equation}
The generator for which the Lagrangian density (\ref{34_eqn}) admits Noether symmetries is given by \cite{Olver}
\begin{equation}\label{generator}
Y=\zeta \frac{\partial }{\partial t}+{{\beta}}^{i}\frac{\partial }{\partial {q}^{i}},
\end{equation}
where $q^i$ are the generalized coordinates in a $d$-dimensional configuration space $Q=\{ q^i,~i=1,...,d \}$. Also it would be worthwhile to mention here that components $\zeta$ and ${\beta}^{i}$ of the above Noether symmetry generator are the multivariable functions of $t$ and $q^i$. For example, in our case $\zeta\equiv\zeta(t,a,\mathcal{R},\varphi,\chi)$ and $q^i\equiv q^i(t,a,\mathcal{R},\varphi,\chi)$, $i=1,...4$.
Also for the existence of Noether symmetry, the Lagrangian (\ref{34_eqn}) must satisfy
\begin{equation}\label{operators}
Y^{[1]}\mathcal{L}+\mathcal{L}(D_t\zeta)=D_t\Psi,
\end{equation}
where the first prolongation $Y^{[1]}$ of the generator (\ref{generator}) is given by
\begin{equation}
Y^{[1]}\equiv Y+\dot{{\beta}}^i\frac{\partial }{\partial {\dot{q}}^{i}},
\end{equation}
where
\begin{itemize}

\item $\Psi\equiv \Psi(t,q^i)$ is known as the Noether gauge function,
\item $D_t$ is the total derivative defined as $D_t\equiv \frac{\partial }{\partial t}+\dot{q}^i \frac{\partial }{\partial q^i}$ and
$\dot{{\beta}}^i\equiv D_t{\beta}^i-\dot{q}^iD_t\zeta$.
\end{itemize}
The most important component of Noether symmetries is the first integral of motion or also known as the conserved quantity. The conserved quantity corresponding to any Noether symmetry generator $Y$ is given by
\begin{equation}\label{integral}
I=-\zeta E_{\mathcal{L}}+{{\beta}}^i\frac{\partial \mathcal{L}}{\partial  {\dot{q}}^{i}}-\Psi.
\end{equation}
It is important to mention here that these conserved quantities play an important role in deriving important cosmological solutions.
Now we get a set of PDEs also known as determining equation after using Eq.(\ref{operators}) and equating the coefficients of $\dot{a}^3,~\dot{a}^2\dot{\mathcal{R}},~\dot{a}^2\dot{\varphi},~\dot{a}^2\dot{\chi}$,
$\dot{\mathcal{R}}^2,~\dot{a}\dot{\varphi},~\dot{a}\dot{\mathcal{R}},~\dot{a}\dot{\chi},~\dot{a}^2,~\dot{\varphi}^2,~\dot{a},
~\dot{\mathcal{R}},~\dot{\varphi},~\dot{\chi},~1$. Thus in our case we get a system of $15$ PDEs
\begin{eqnarray}\label{determining1}
\zeta_{a}=0,~~~~~~~~~~~\zeta_{\mathcal{R}}=0,~~~~~~~~~~~\zeta_{\varphi}=0,~~~~~~~~\zeta_{\chi}=0,~~~~~~~~f_{\mathcal{R}\mathcal{R}}{\beta^1},_{\mathcal{R}}=0,
\end{eqnarray}
\begin{eqnarray}\label{determining2}
6a^2f_{\mathcal{R}\mathcal{R}}{\beta^1},_{t}+\Psi,_{\mathcal{R}}=0, ~~~~~~~~ \epsilon f_{\chi}a^3{\beta^3},_{t}-\Psi,_{\varphi}=0, ~~~~~~~~6a^2f_{\mathcal{R}\chi}{\beta^1},_{t}+\Psi,_{\chi}=0,
\end{eqnarray}
\begin{eqnarray}\label{determining3}
3f_{\chi}{\beta^1}+a(f_{\mathcal{R}\chi}{\beta^2}+f_{\chi\varphi}{\beta^3}+f_{\chi\chi}{\beta^4})+af_{\chi}\zeta,_{t}=0,~~~
12af_{\mathcal{R}}{\beta^1},_{t}+6a^2(f_{\mathcal{R}\mathcal{R}}{\beta^2},_{t}+f_{\mathcal{R}\varphi}{\beta^3},_{t}+f_{\mathcal{R}\chi}{\beta^4},_{t})+\Psi,_{a}=0,
\end{eqnarray}
\begin{eqnarray}\label{determining4}
f_{\mathcal{R}}{\beta^1}+a(f_{\mathcal{R}\mathcal{R}}{\beta^2}+f_{\mathcal{R}\varphi}{\beta^3}+f_{\mathcal{R}\chi}{\beta^4})+a^2(f_{\mathcal{R}\mathcal{R}}{\beta^2},_{a}+f_{\mathcal{R}\varphi}{\beta^3},_{a}+f_{\mathcal{R}\chi}{\beta^4},_{a})
+2af_{\mathcal{R}}{\beta^1},_{a}-af_{\mathcal{R}}\zeta,_{t}=0,
\end{eqnarray}
\begin{eqnarray}\label{determining5}
2f_{\mathcal{R}\varphi}{\beta^1}+a(f_{\mathcal{R}\mathcal{R}\varphi}{\beta^2}+f_{\mathcal{R}\varphi\varphi}{\beta^3}+f_{\mathcal{R}\varphi\chi}{\beta^4})+a(f_{\mathcal{R}\varphi}{\beta^1},_{a}+f_{\mathcal{R}\mathcal{R}}{\beta^2},_{\varphi}+f_{\mathcal{R}\varphi}{\beta^3},_{\varphi}
+f_{\mathcal{R}\chi}{\beta^4},_{\varphi})+2f_{\mathcal{R}}{\beta^1},_{\varphi}=0,
\end{eqnarray}
\begin{eqnarray}\label{determining6}
2f_{\mathcal{R}\mathcal{R}}{\beta^1}+a(f_{\mathcal{R}\mathcal{R}\mathcal{R}}{\beta^2}+f_{\mathcal{R}\mathcal{R}\varphi}{\beta^3}+
f_{\mathcal{R}\mathcal{R}\chi}{\beta^4})+
a(f_{\mathcal{R}\mathcal{R}}{\beta^1},_{a}+f_{\mathcal{R}\mathcal{R}}{\beta^2},_{\mathcal{R}}+f_{\mathcal{R}\varphi}{\beta^3},_{\mathcal{R}}+f_{\mathcal{R}\chi}{\beta^4},_{\mathcal{R}})+2f_{\mathcal{R}}{\beta^1},_{\mathcal{R}}-af_{\mathcal{R}\mathcal{R}}\zeta,_{t}=0,
\end{eqnarray}
\begin{eqnarray}\label{determining7}
2f_{\mathcal{R}\chi}{\beta^1}+a(f_{\mathcal{R}\mathcal{R}\chi}{\beta^2}+f_{\mathcal{R}\varphi\chi}{\beta^3}+f_{\mathcal{R}\chi\chi}{\beta^4})+a(f_{\mathcal{R}\chi}{\beta^1},_{a}+
f_{\mathcal{R}\mathcal{R}}{\beta^2},_{\chi}+f_{\mathcal{R}\varphi}{\beta^3},_{\chi}+f_{\mathcal{R}{\beta^2}}{\beta^4},_{\chi})+2f_{\mathcal{R}}{\beta^1},_{\chi}=0,
\end{eqnarray}
\begin{eqnarray}\nonumber
&&(3a^2{\beta^1}+a^3\zeta,_{t})(f-\mathcal{R}f_{\mathcal{R}}-\chi f_{\chi})+a^3[(-\mathcal{R}f_{\mathcal{R}\mathcal{R}}-\chi f_{\chi \mathcal{R}}){\beta^2}+(f_{\varphi}-\mathcal{R}f_{\mathcal{R}\varphi}-\chi f_{\chi \varphi}){\beta^3}+
(-\mathcal{R}f_{\mathcal{R}\chi}-\chi f_{\chi \chi}){\beta^4}]-\Psi,_{t}=0.\\\label{determining}
&&
\end{eqnarray}
In the next section, we provide a comprehensive analysis by solving this system of PDEs for different cases.

\section{Conserved Quantities}

In this section, we solve the system of PDEs (\ref{determining1})-(\ref{determining}) to get the Noether symmetries $Y=\zeta \partial_t +{\beta}^i\partial_i$. Since the above system depends on the function $f(\mathcal{R},\varphi,\chi)$ along with some other unknowns, so it is difficult to find a simultaneous solution without assigning some appropriate values to $f(\mathcal{R},\varphi,\chi)$. However, if we see first equation of the the system (\ref{determining}), we come across with at least a trivial symmetry $Y=\partial_t$ independent of the choice of any specific $f(\mathcal{R},\varphi,\chi)$ gravity model. Thus, we start our analysis by choosing different cosmological models.

\subsection{$f(\mathcal{R})$ Gravity}

This is an important test case which enables us to revisit usual metric $f(\mathcal{R})$ gravity. It may be noticed from the last equation of (\ref{determining1}) that either $f_{\mathcal{R}\mathcal{R}}=0$ or ${\beta^1},_{\mathcal{R}}=0$. So if we assume $f_{\mathcal{R}\mathcal{R}}=0$ and ${\beta^1},_{\mathcal{R}}\neq0$, then Eq.(\ref{determining6}) provides $f_{\mathcal{R}}=0$. In this situation both $\zeta$ and $\Psi$ becomes functions dependent on $t$ only and Eq. (\ref{determining}) simplifies to
\begin{eqnarray}\label{1000}
(3a^2{\beta^1}+a^3\zeta,_{t})f-\Psi,_{t}=0.
\end{eqnarray}
Differentiating Eq.(\ref{1000}) with respect to $\mathcal{R}$ yields ${\beta}^{1},_{\mathcal{R}}=0$, which is contrary to our supposition. Thus $f(\mathcal{R})$ gravity is constrained with the condition that $f_{\mathcal{R}\mathcal{R}}\neq0$ \cite{22}. So for the sake of simplicity, we consider $f(\mathcal{R},\varphi,\chi)=f_0\mathcal{R}^{\frac{3}{2}}$. It is worthwhile to mention here that this $f(\mathcal{R})$ model has already been used in literature \cite{22, me3}. Manipulating Eqs.(\ref{determining1}-\ref{determining}), we get
\begin{eqnarray}\label{(Sol)}
\zeta=c_1t+c_2,~~~~~~~{\beta^1}&=&\frac{2c_1a^2+3c_3t+3c_4}{3a},~~~~~{\beta^2}=-2\mathcal{R}\frac{c_1a^2+c_3t+c_4}{a^2},\\\Psi&=&-9c_3a\sqrt{\mathcal{R}}+c_5, ~~~~{\beta}^3={\beta}^4=0.
\end{eqnarray}
The Noether symmetry generators turn out to be
\begin{eqnarray}\label{(NGg)}
Y_1=t\frac{\partial}{\partial
t}+\frac{2}{3}a\frac{\partial}{\partial
a}-2\mathcal{R}\frac{\partial}{\partial \mathcal{R}},~~~Y_2=\frac{\partial}{\partial
t},~~~Y_3=ta^{-1}\frac{\partial}{\partial
a}-2\mathcal{R}ta^{-2}\frac{\partial}{\partial
\mathcal{R}},~~~Y_4=a^{-1}\frac{\partial}{\partial
a}-2\mathcal{R}a^{-2}\frac{\partial}{\partial \mathcal{R}}.
\end{eqnarray}
It is important to mention here that these are exactly the same as obtained in \cite{me3}.
\begin{figure}
\centering \epsfig{file=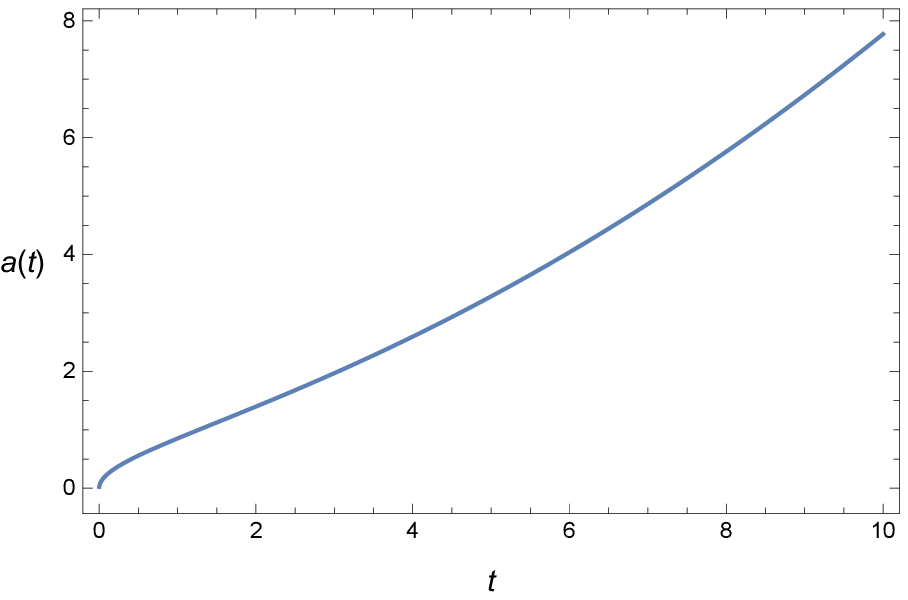, width=.45\linewidth,
height=1.9in}~~~~~\epsfig{file=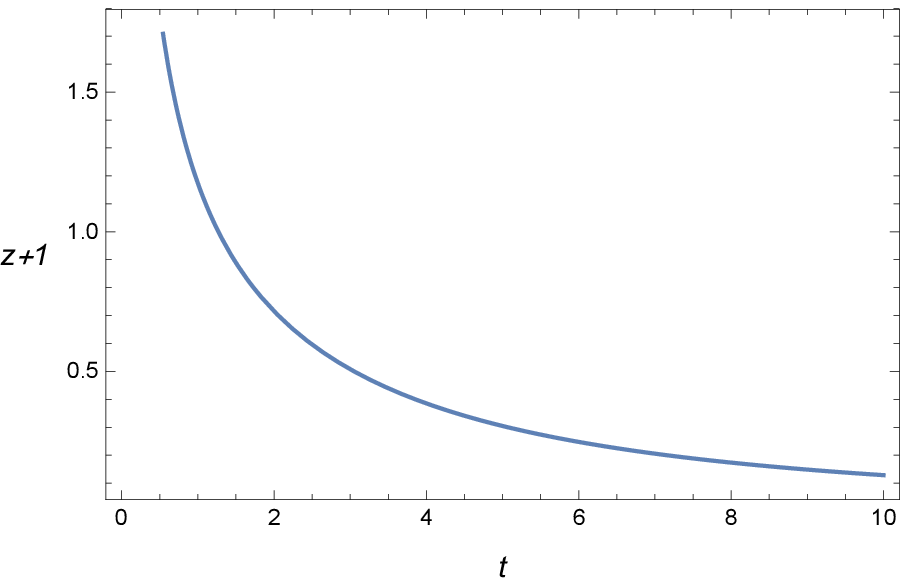, width=.45\linewidth,
height=1.9in}\caption{\label{Fig.1} Evolution of scale factor versus cosmic time and behaviour of redshift function for the case $f(\mathcal{R},\varphi,\chi)=f_0\mathcal{R}^{\frac{3}{2}}$}
\end{figure}
Moreover, using Eq.(\ref{integral}) first integrals also known as conserved quantities turn out to be
\begin{equation}\nonumber
I_{1}=9a{\dot{a}}^2f_0\mathcal{R}^{\frac{1}{2}}-\frac{1}{2}a^3f_0\mathcal{R}^{\frac{3}{2}}+
\frac{9}{2}a^2\dot{a}\dot{\mathcal{R}}f_0\mathcal{R}^{\frac{-1}{2}},~~~
I_{2}=9at{\dot{a}}^2f_0\mathcal{R}^{\frac{1}{2}}-\frac{1}{2}a^3tf_0\mathcal{R}^{\frac{3}{2}}+
\frac{9}{2}a^2t\dot{a}\dot{\mathcal{R}}f_0\mathcal{R}^{\frac{-1}{2}}-3a^2\dot{a}f_0\mathcal{R}^{\frac{1}{2}}-3a^3\dot{\mathcal{R}}f_0\mathcal{R}^{\frac{-1}{2}},
\end{equation}
\begin{eqnarray}\label{1001}
I_{3}=9af_0\mathcal{R}^{\frac{1}{2}}-9t\dot{a}f_0\mathcal{R}^{\frac{1}{2}}-
\frac{9}{2}ta\dot{\mathcal{R}}f_0\mathcal{R}^{\frac{-1}{2}},~~~~~~~~~~~
I_{4}=-9\dot{a}f_0\mathcal{R}^{\frac{1}{2}}-
\frac{9}{2}a\dot{\mathcal{R}}f_0\mathcal{R}^{\frac{-1}{2}}.
\end{eqnarray}
It is mentioned here these conserved quantities play an important role in deriving important cosmological solutions though after some tedious calculations. As an example, we present here one important solution. Last integral of Eq.(\ref{1001}) can be rearranged as
\begin{equation}\label{1245}
\dot{a}+\frac{1}{2}a\frac{\dot{\mathcal{R}}}{\mathcal{R}}+\frac{I_4}{9f_0\mathcal{R}^{\frac{1}{2}}}=0.
\end{equation}
This is an interesting differential equation and one can try to find both numerical and analytical solutions if possible to investigate the evolution of universe. After using the initial condition $a(0)=0$ and considering the value of parameters in such a way that $\frac{I_4}{f_0}=2$, a numerical solution is obtained. It is evident from Fig. 1 that the behavior scale factor is physical due to increasing trend.
As far as analytic approach is concerned, Eq.(\ref{1245}) provides a solution of the form
\begin{equation}
a=\mathcal{R}^{-{\frac{1}{2}}}\big[a_0-\frac{I_4t}{9f_0}\big].
\end{equation}
This differential equation after using the value of Ricci scalar from Eq.(\ref{Ricci}) yields an interesting exact solution for the cosmic scale factor
\begin{equation}\label{1231}
a(t)=(c_4t^4+c_3t^3+c_2t^2+c_1t+c_0)^{\frac{1}{2}},
\end{equation}
such that
\begin{equation}
c_4=\frac{{I_4}^2}{2916{f_0}^2},~~~~c_3=-\frac{a_0I_4}{81f_0},~~~~c_2=\frac{{a_0}^2}{6},~~~~c_1=a_1-\frac{a_0f_0}{I_4},~~~c_0=a_2+\frac{9{a_0}^4{f_0}^2}{4{I_4}^2},
\end{equation}
where ${a_i}'s$ are integration constants. It is worthy to mention here that similar solutions have already been reported in the context of $f(\mathcal{R})$ gravity \cite{22a,22b}. The values of constants ${c_i}'s$ give the information about the cosmological evolution. For instance, $c_4\neq 0$ provides a power law inflation while, a radiation dominated stage is obtained if the regime is dominated by the linear term in $c_1$ \cite{22a}.

Now manipulating first two conserved quantities of Eq.(\ref{1001}), we get
\begin{equation}
\dot{a}a^2+a^3\frac{\dot{\mathcal{R}}}{\mathcal{R}}-\frac{I_1t-I_2}{3f_0\mathcal{R}^{\frac{1}{2}}}=0,
\end{equation}
whose solution for scale factor $a$ is obtained as
\begin{equation}\label{201}
a=\mathcal{R}^{-1}\big[a_3+\frac{1}{f_0\mathcal{R}^3}\int(I_1t-I_2)\mathcal{R}^{\frac{5}{2}}~dt\big]^{\frac{1}{3}}.
\end{equation}
The integral involved in this equation can be solved once we assume any suitable functional form for $\mathcal{R}$. A simplest choice can be
\begin{equation}
\mathcal{R}=\mathcal{R}_0t^r,~~~~~~r\neq 0,~1,~~~~\mathcal{R}_0\neq 0,~~~~r,\mathcal{R}_0~\in \mathfrak{R}.
\end{equation}
In this case, the differential equation (\ref{201}) provides a cosmological solution
\begin{equation}\label{20212}
a(t)=\frac{1}{\mathcal{R}_0t^r}\big[a_3+\frac{2^{\frac{1}{3}}}{f_0{\mathcal{R}_0}^{\frac{1}{2}}t^{\frac{r-2}{2}}}\big(\frac{I_1t}{5r+4}-\frac{I_2}{5r+2}\big)\big]^{\frac{1}{3}},~~~~~~r\neq -\frac{2}{5},~-\frac{4}{5}.
\end{equation}
To our best of knowledge, this is an important new solution in the context of $f(\mathcal{R})$ gravity. It is interesting to notice that one can obtain many cosmic solutions in modified $f(\mathcal{R})$ gravity which is due to the consequence of Eq.(\ref{502}). Moreover, different scenarios of cosmic evolution can be discussed using Eq.(\ref{20212}). For example, when $r=\frac{2}{5}$, the scale factor diverges as $t\rightarrow 0$ indicating the existence of Big Rip singularity \cite{bigrip, bigrip1}.

\subsection{$f(\mathcal{R},\varphi,\chi)$ Gravity}

In this section, we investigate the conserved quantities for a broader class. For this purpose, we consider the following cases:\\\\ $\mathbf{Case(i)}$: $f(\mathcal{R},\varphi,\chi)=\varphi \mathcal{R}^n,~~~n\neq 0,1.$ \\\\
In this case, manipulation of Eqs.(\ref{determining1}-\ref{determining}) yields,
\begin{eqnarray}\label{(Sol3)}
\zeta&=&c_1t+c_2,~~~~~~~{\beta^1}=c_1a \ln\varphi+c_3a^{-1}+c_4a,~~~~~{\beta^2}=-2\mathcal{R}c_1-2\mathcal{R}a^{-2}c_3-\frac{\mathcal{R}^{2-n}a^{-1}}{n\varphi }c_5,\\{\beta^3}&=&\varphi(2n-1-3\ln\varphi)c_1+\varphi(2n-3)a^{-2}c_3-3\varphi c_4+\mathcal{R}^{1-n}a^{-1}c_5, ~~~~\Psi={\beta}^4=0.
\end{eqnarray}
Here the Noether symmetry generators take the form
\begin{eqnarray}\nonumber
Y_1=t\frac{\partial}{\partial
t}+a\ln\varphi\frac{\partial}{\partial
a}-2\mathcal{R}\frac{\partial}{\partial \mathcal{R}}&+&\varphi(2n-1-3\ln\varphi)\frac{\partial}{\partial
\varphi},~~~Y_2=\frac{\partial}{\partial
t},~~~Y_3=a^{-1}\frac{\partial}{\partial
a}-2\mathcal{R}a^{-2}\frac{\partial}{\partial
\mathcal{R}}+\varphi(2n-3)a^{-2}\frac{\partial}{\partial
\varphi},\\\label{(NGg)}
Y_4&=&a\frac{\partial}{\partial
a}-3\varphi\frac{\partial}{\partial \varphi},~~~~~~~~~~~Y_5=-\frac{\mathcal{R}^{2-n}a^{-1}}{n\varphi }\frac{\partial}{\partial
\mathcal{R}}+\mathcal{R}^{1-n}a^{-1}\frac{\partial}{\partial
\varphi}.
\end{eqnarray}
First integrals become
\begin{equation}\nonumber
I_{1}=t[6a{\dot{a}}^2f_0n\varphi \mathcal{R}^{n-1}
6a^2\dot{a}\dot{\mathcal{R}}f_0n\varphi(n-1)\mathcal{R}^{n-2}-a^3(n-1)f_0\varphi \mathcal{R}^{n}]-6f_0na^2\ln\varphi \mathcal{R}^{n-2}[2\dot{a}\mathcal{R}+a\dot{\mathcal{R}}(n-1)]+12f_0n(n-1)\varphi \mathcal{R}^{n-1}a^2\dot{a},
\end{equation}
\begin{eqnarray}\nonumber
I_{2}=6a{\dot{a}}^2f_0n\varphi \mathcal{R}^{n-1}
6a^2\dot{a}\dot{\mathcal{R}}f_0n\varphi(n-1)\mathcal{R}^{n-2}-a^3(n-1)f_0\varphi \mathcal{R}^{n},~~~
I_{3}=12\dot{a}f_0n(n-2)\varphi \mathcal{R}^{n-1}-6af_0n(n-1)\varphi \mathcal{R}^{n-2}\dot{\mathcal{R}}.
\end{eqnarray}
\begin{eqnarray}\label{120001}
I_{4}=-12f_0na^2\dot{a}\varphi \mathcal{R}^{n-1}-6f_0n(n-1)a^3\varphi \mathcal{R}^{n-2}\dot{\mathcal{R}},~~~~~~~~~~~I_{5}=6f_0(n-1)a\dot{a}.
\end{eqnarray}
Here we can also construct many important cosmological solutions. Last integral of Eq.(\ref{10001}) is much easier to deal with and give a straight away solution of the scale factor $a=[\frac{2I_5}{6f_0(n-1)}t+a_4]^{1/2}$. Moreover, third conserved quantity can be used to construct the solution for scale factor in the form
\begin{equation}\label{20019}
a=\mathcal{R}^{{\frac{n-1}{2(n-2)}}}\big[a_5+\frac{I_3}{12f_0n(n-2)}\int\frac{\mathcal{R}^{\frac{(1-n)(2n-3)}{2(n-2)}}}{\varphi}dt\big].
\end{equation}
It is interesting to notice that when $\varphi=1$, this differential equation yields the same solution (\ref{1231}) in $f(\mathcal{R})$ theory of gravity when $n={\frac{3}{2}}$ and the case reduces to ordinary GR when $n=1$. However, some interesting solutions can be developed by choosing appropriate values for the scalar field $\varphi$. Similarly, manipulating fourth conserved quantity of Eq.(\ref{120001}), we get
\begin{equation}\label{20018}
a=\mathcal{R}^{\frac{1-n}{2}}\big[a_6-\frac{I_4}{4f_0n}\int \frac{\mathcal{R}^{\frac{n-1}{2}}}{\varphi}~dt\big]^{\frac{1}{3}}.
\end{equation}
The integral involved in both the equations (\ref{20019},\ref{20018}) can be easily solved once we assume some suitable functional form for $\mathcal{R}$ and $\varphi$.\\\\
$\mathbf{Case(ii)}$: $f(\mathcal{R},\varphi,\chi)=\varphi^m \mathcal{R}^{\frac{19}{14}},~~~m\neq 0,1.$ \\\\
In this case, a simultaneous solution of Eqs.(\ref{determining1}-\ref{determining}) gives,
\begin{eqnarray}\label{(Sol3)}
\zeta&=&c_1t+c_2,~~~~~~~{\beta^1}=c_1ma \ln\varphi+c_3a+c_4a^{-1},~~~~~{\beta^2}=-2\mathcal{R}c_1-2\mathcal{R}a^{-2}c_4-\frac{14}{19}\mathcal{R}^{\frac{9}{14}}c_5a^{-1}m{\varphi}^{-m},\\{\beta^3}&=&-3\varphi\big(\ln\varphi-\frac{4}{7m}\big) c_1-\frac{3\varphi}{m}c_3-\frac{6\varphi}{21m}a^{-2}c_4+\varphi^{1-m}\mathcal{R}^{-\frac{5}{14}}a^{-1}c_5, ~~~~~~~\Psi={\beta^1}^4=0.
\end{eqnarray}
\begin{figure}
\centering \epsfig{file=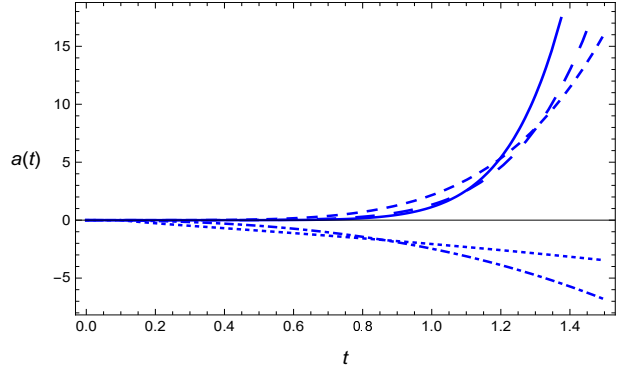, width=.45\linewidth,
height=1.9in}\caption{\label{Fig.1} Evolution of scale factor factor for $f(\mathcal{R},\varphi,\chi)=\varphi^m \mathcal{R}^{\frac{19}{14}}$ ; $m=2.5$ (Solid), $m=2$ (Dashed), $m=1.5$ (Small Dashed), $m=-0.125$ (Dotted), $m=0.5$ (Dot Dashed)}
\end{figure}
Here the Noether symmetry generators take the form
\begin{eqnarray}\nonumber
Y_1&=&t\frac{\partial}{\partial
t}+ma\ln\varphi\frac{\partial}{\partial
a}-2\mathcal{R}\frac{\partial}{\partial \mathcal{R}}+3\varphi\big(\frac{4}{7m}-\ln\varphi\big)\frac{\partial}{\partial
\varphi},~~~~~~~~Y_2=\frac{\partial}{\partial
t},~~~~~~~~~Y_3=a\frac{\partial}{\partial
a}-\frac{3\varphi}{m}\frac{\partial}{\partial \varphi},\\\label{(NGg)}
Y_4&=&a^{-1}\frac{\partial}{\partial
a}-2\mathcal{R}a^{-2}\frac{\partial}{\partial \mathcal{R}}-\frac{6\varphi}{21m}a^{-2}\frac{\partial}{\partial \varphi},
~~~~~Y_5=-\frac{14}{19}\mathcal{R}^{\frac{9}{14}}a^{-1}m{\varphi}^{-m}\frac{\partial}{\partial
\mathcal{R}}++\varphi^{1-m}\mathcal{R}^{-\frac{5}{14}}a^{-1}\frac{\partial}{\partial
\varphi}.
\end{eqnarray}
First integrals turn out to be
\begin{equation}\nonumber
I_{1}=t[\frac{57}{7}a{\dot{a}}^2f_0\varphi^m \mathcal{R}^{\frac{5}{14}}+
\frac{285}{98}a^2\dot{a}\dot{\mathcal{R}}f_0\varphi^m\mathcal{R}^{-\frac{9}{14}}-\frac{5}{14}a^3f_0\varphi^m \mathcal{R}^{\frac{9}{14}}]
-6f_0ma^2\varphi^m\ln\varphi [\frac{19}{7}\dot{a}\mathcal{R}^{\frac{5}{14}}+\frac{95}{196}a\dot{\mathcal{R}}\mathcal{R}^{-\frac{9}{14}}]+\frac{285}{49}f_0\varphi^m \mathcal{R}^{\frac{5}{14}}a^2\dot{a},
\end{equation}
\begin{eqnarray}\nonumber
I_{2}=\frac{57}{7}a{\dot{a}}^2f_0\varphi^m \mathcal{R}^{\frac{5}{14}}+
\frac{285}{98}a^2\dot{a}\dot{\mathcal{R}}f_0\varphi^m\mathcal{R}^{-\frac{9}{14}}-\frac{5}{14}a^3f_0\varphi^m \mathcal{R}^{\frac{9}{14}},~~~
I_{3}=-6f_0a^2\varphi^m[\frac{19}{7}\dot{a}\mathcal{R}^{\frac{5}{14}}+\frac{95}{196}a\dot{\mathcal{R}}\mathcal{R}^{-\frac{9}{14}}].
\end{eqnarray}
\begin{eqnarray}\label{10001}
I_{4}=-6f_0\varphi^m[\frac{19}{7}\dot{a}\mathcal{R}^{\frac{5}{14}}+\frac{95}{196}a\dot{\mathcal{R}}\mathcal{R}^{-\frac{9}{14}}]
+\frac{285}{49}f_0\varphi^m \mathcal{R}^{\frac{5}{14}}\dot{a},~~~~~~~~~~~I_{5}=\frac{15}{7}f_0ma\dot{a}.
\end{eqnarray}
Here we also get five conserved quantities as in the previous case. It is due to the fact that both cases are similar in nature, the only difference is that in first case Ricci power law model is used while in this case scalar field power law form is used for a better comparative analysis. Here we can also reconstruct many important cosmological solutions using these conserved quantities. As in the last case, fifth integral of Eq.(\ref{10001}) provides a solution $a=[\frac{14I_5}{15f_0m}t+a_7]^{1/2}$. The fourth conserved quantity can be used to construct the solution for scale factor in the form
\begin{equation}\label{2019}
a=\mathcal{R}^{-{\frac{5}{18}}}\big[a_8-\frac{49I_4}{513f_0}\int\frac{\mathcal{R}^{-\frac{5}{63}}}{\varphi^m}dt\big].
\end{equation}
It would be worthwhile to mention here that the scalar field parameter $m$ can play an important role in the evolution of universe. For example, when we assume scalar field as a quadratic function of cosmic time and $\mathcal{R}=\mathcal{R}_0t^{-\frac{63m}{5}}$, the evolution of scale factor becomes interesting. It can be seen from Fig. $2$ that for $m>1$, the scale factor remains positive and shows an increasing behavior. In fact, an accelerated expansion phase is observed as $m$ increases. Also, when $m<1$, the scale factor shows a negative and decreasing behavior. Thus with the considered parametric values involved in the analysis, the scalar field becomes very important when $m>1$.\\\\
Similarly, manipulating third conserved quantity of Eq.(\ref{10001}), we get
\begin{equation}\label{2018}
a=\mathcal{R}^{-\frac{5}{28}}\big[a_9-\frac{7I_3}{38f_0}\int \frac{\mathcal{R}^{\frac{5}{28}}}{\varphi^m}~dt\big]^{\frac{1}{3}}.
\end{equation}
A similar numerical analysis can also be done here by solving the integral involved in Eq.(\ref{2018}) if we consider some suitable forms for $\mathcal{R}$ and $\varphi$. \\\\
$\mathbf{Case(iii)}$: $f(\mathcal{R},\varphi,\chi)=\varphi \mathcal{R}+h(\chi)$ \\\\
Here we discuss the case with more general form of $f(\mathcal{R},\varphi,\chi)$ model. For, this purpose we consider $f(\mathcal{R},\varphi,\chi)=\varphi \mathcal{R}+h(\chi)$, where $h(\chi)$ is an arbitrary analytic function of $\chi$.
Solving Eqs.(\ref{determining1}-\ref{determining}) simultaneously, it follows
\begin{eqnarray}\label{(Sol3)}
\zeta&=&c_1,~~~~{\beta^3}=c_2a^{-1},~~~\Psi=c_3,~~~~{\beta^1}={\beta^2}=\beta^4=0.
\end{eqnarray}
Here the Noether symmetry generators become
\begin{eqnarray}
Y_1=\frac{\partial}{\partial
t},~~~~~~~Y_2=a^{-1}\frac{\partial}{\partial
\varphi}.
\end{eqnarray}
\begin{figure}
\centering \epsfig{file=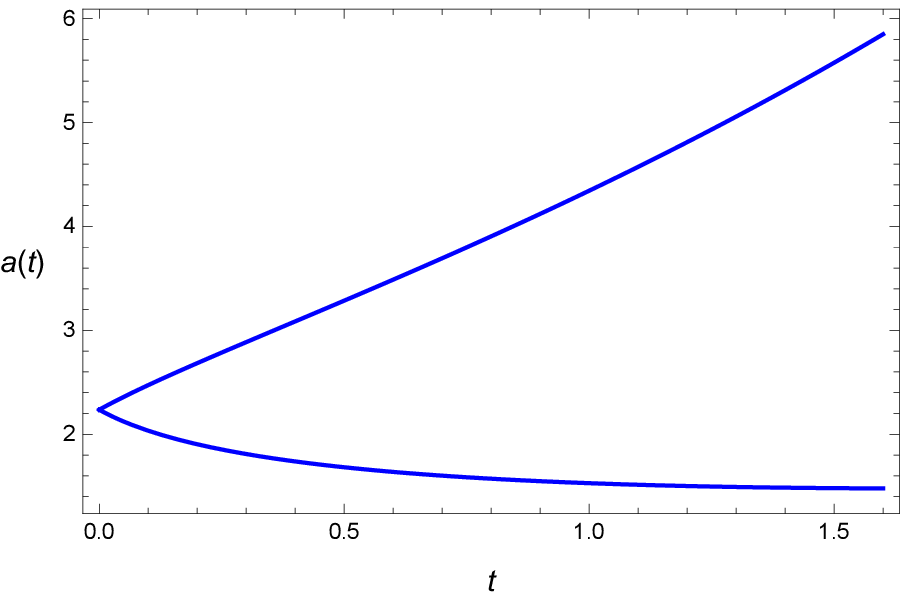, width=.45\linewidth,
height=1.9in}~~~~~\epsfig{file=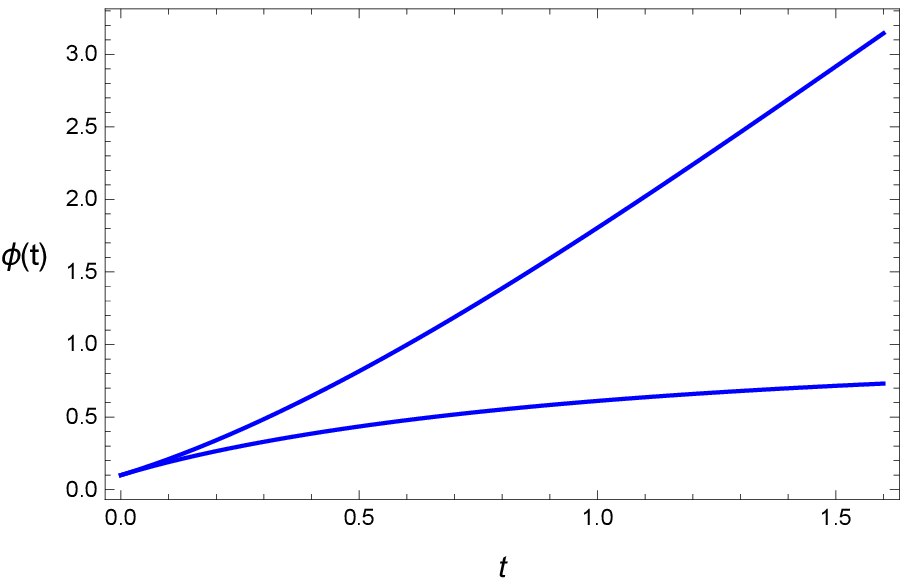, width=.45\linewidth,
height=1.9in}\caption{\label{Fig.1} Evolution of scale factor and canonical scalar field ($\epsilon=1$) for $f(\mathcal{R},\varphi,\chi)=\varphi \mathcal{R}+h(\chi)$ with $h(\chi)=k_1\chi+k_2$}
\end{figure}
Many cosmic solutions can be constructed for different choices of the function $h(\chi)$. However for the present analysis, we consider the simplest case $h(\chi)=k_1\chi+k_2$, where $k_1\neq 0$ and $k_2$ are arbitrary real coefficients.
Corresponding conserved quantities in this case turn out to be
\begin{equation}\label{(Sol33)}
I_{1}=6a{\dot{a}}^2\varphi+a^3\big(k_2+\frac{1}{2}k_1\epsilon{\dot{\varphi}}^{2}\big),~~~~~~~~
I_{2}=\epsilon a^2\dot{\varphi}.
\end{equation}
The second conserved quantity suggests that the growth of scale factor depends on the scalar field and varies with inverse proportion. Moreover, the simultaneous solution of both first integrals in Eq.(\ref{(Sol33)}) provide
\begin{equation}\label{(Sol303)}
6\varphi\big[{\frac{I_2}{\epsilon \dot{\varphi}}}\big]^{\frac{1}{2}}\bigg[\frac{d}{dt}\big({\frac{I_2}{\epsilon \dot{\varphi}}}\big)^{1/2}\bigg]^2
+\big[{\frac{I_2}{\epsilon \dot{\varphi}}}\big]^{\frac{3}{2}}\big[k_2+\frac{1}{2}k_1\epsilon{\dot{\varphi}}^{2}\big]-I_1=0.
\end{equation}
One can try for the possible solution for the scalar field $\varphi$ with suitable initial conditions. As an example, a numerical solution is obtained for canonical scalar field $\epsilon=1$. Two solution curves in Fig. $3$ are due to the square root term involved. The evolution of scale factor with blue colored curve seems more physical due to increasing behavior.

\section{Outlook}

Noether symmetries are not simply a mechanism to deal with the dynamical solutions, but also their possible existence may provide some feasible conditions so that one can choose some viable universe models according to recent observations. Lagrangian multipliers are useful to re-shape the Lagrangian into its canonical form which may prove to be quite useful to reduce the dynamics of the system and eventually help in determining the exact solutions.
This paper is devoted to provide a detailed discussion about the Noether symmetries of the flat FRW universe model with in $f(\mathcal{R},\varphi,\chi)$ gravity set up.
We have calculated the Lagrangian for FRW universe with in $f(\mathcal{R},\varphi,\chi)$ theory background. The existence of Noether symmetries and corresponding conserved quantities is considered important in the literature and plays an important role to explore the exact solutions of field equations. A brief summary of results of this paper is as follows:
\begin{itemize}
\item The exact solutions of Noether equations have been divided mainly in two parts. The first case is focussed on revisiting usual metric $f(\mathcal{R})$ gravity.
It is worthwhile to mention here that the $f(\mathcal{R})$ gravity is constrained with the condition that $f_{\mathcal{R}\mathcal{R}}\neq0$. Thus for the sake of simplicity, we have considered $f(\mathcal{R},\varphi,\chi)=f_0\mathcal{R}^{\frac{3}{2}}$, where $f_0\neq 0$ is a real arbitrary parameter. The obtained Noether symmetries turn out to be same as already available in literature \cite{22, me3}. However, we use conserved quantities to investigate both numerical and analytical solutions to study the evolution of universe. After using some appropriate initial condition and suitable values of parameters involved, a numerical solution is obtained. It is evident from the behavior scale factor as shown in Fig. $1$ that universe expanded at an early time with deceleration phase and then later on accelerated cosmic expansion could be observed. Analytical approach provides an exact solution that have already been suggested to exist in the context of $f(\mathcal{R})$ gravity \cite{22a,22b}. We also report a new exact solution (\ref{20212}) in the context of $f(\mathcal{R})$ gravity which for a special case indicates the existence of Big Rip singularity.
\item The second part deals with the more general form of $f(\mathcal{R},\varphi, \chi)$ gravity model. Three subcases have been discussed under this category. The first case provides the exact solutions of FRW universe when $f(\mathcal{R},\varphi,\chi)=\varphi \mathcal{R}^n,~~~n\neq 0,1$. Five conserved quantities have been obtained in this case. Three of them have been used to provide a new class of exact solutions in the context of $f(\mathcal{R},\varphi, \chi)$ gravity.
\item We further consider $f(\mathcal{R},\varphi,\chi)=\varphi^m \mathcal{R}^{\frac{19}{14}}$ for finding the Noether symmetry generators and corresponding first integrals.  Here we also get five conserved quantities as in the previous case. It is due to the fact that both cases are similar in nature, the only difference is that in first case Ricci power law model is used while in this case scalar field power law form is used for a better comparative analysis. Many solutions are possible using conserved quantities in this case, however, one solution has been reported for the discussion. It is worth mentioning that the scalar field parameter $m$ plays an important role in the evolution of universe. For example, when scalar field is assumed as a quadratic function of cosmic time and $\mathcal{R}=\mathcal{R}_0t^{-\frac{63m}{5}}$, the evolution of scale factor becomes interesting. It can be seen from Fig. $2$ that for $m>1$, the scale factor remains positive and shows an increasing behavior. In fact, an accelerated expansion phase is observed as $m$ increases. Also, when $m<1$, the scale factor shows a negative and decreasing behavior. Thus with the considered parametric values involved in this analysis, the scalar field becomes very important when $m>1$.
\item The last cases provides the Noether symmetries for $f(\mathcal{R},\varphi,\chi)=\varphi \mathcal{R}+h(\chi)$. Here we obtain two Noether symmetry generators after solving the determining equations simultaneously. Many cosmic solutions can be constructed here for different choices of the function $h(\chi)$. However for the present analysis, we consider the simplest case $h(\chi)=k_1\chi+k_2$, where $k_1\neq 0$ and $k_2$ are arbitrary real coefficients. A conserved quantity in this case suggests that the growth of scale factor depends on the scalar field and varies with inverse proportion. A non-linear differential equation has been formed in terms of scalar field $\varphi$. We can try for the possible solution for the scalar field with some suitable initial conditions. A numerical solution is reported for canonical scalar field $\epsilon=1$. The evolution of scale factor with blue colored curve as shown in Fig. $3$ seems more physical due to increasing behavior.
 \end{itemize}

In nutshell, many other cosmologically viable solutions can be constructed for some specific choice of $f(\mathcal{R},\varphi,\chi)$ gravity models.
A detailed analysis including matter source and with anisotropic background is under process.\newpage

\end{document}